\documentstyle[preprint,aps]{revtex}
\tighten
\begin{document}
\preprint{PITT-09-95, CMU-95-03, DOE-ER/40682-92, LPTHE-95-18, UPRF-95-420}
\draft
\title{\bf Reheating the Post Inflationary Universe}
\author{{\bf D. Boyanovsky$^{(a)}$,
M. D'Attanasio\footnote{Della Riccia fellow}$^{(b)(d)}$,
H. J. de Vega$^{(b)}$,  R. Holman$^{(c)}$,}\\ {\bf
D.-S. Lee$^{(a)}$, A. Singh$^{(c)}$}}
\address
{ (a)  Department of Physics and Astronomy, University of
Pittsburgh, Pittsburgh, PA. 15260, U.S.A. \\
 (b) Laboratoire de Physique Th\'eorique et Hautes Energies,
Universit\'e Pierre et Marie Curie (Paris VI)
et  Universit\'e Denis Diderot (Paris VII),
Tour 16, 1er. \'etage, 4, Place Jussieu
75252 Paris, Cedex 05, FRANCE. Laboratoire Associ\'{e} au CNRS URA280.\\
 (c) Department of Physics, Carnegie Mellon University, Pittsburgh,
PA. 15213, U. S. A. \\
 (d) I.N.F.N., Gruppo Collegato di Parma, ITALIA }
\date{April 1995}
\maketitle
\begin{abstract}

We consider the non-equilibrium evolution of the inflaton field coupled to both
lighter scalars and fermions. The dissipational dynamics of this field is
studied and found to be quite different than that believed in inflationary
models. In particular, the damping time scale for the expectation value of the
zero
 momentum mode of the inflaton  can be
much shorter than that given by the single particle decay rate when
the inflaton amplitudes are large, as in chaotic inflation scenarios. We
find that the reheating temperature may depart considerably from the usual
estimates.
\end{abstract}
\narrowtext
The Universe after inflation is a cold and dark place. The inflationary
expansion phase has driven the matter and radiation energy densities down to
almost zero. If we are to recover the Universe of the standard hot big bang
model, some source of energy (or perhaps more accurately, entropy) must be
enlisted to rethermalize the Universe.

Once inflation has ended,
the standard picture of inflation\cite{linde,abbott} has it that the
zero momentum mode of the inflaton field oscillates about the minimum
of its (effective)  potential. As it
oscillates, the couplings of the inflaton to lighter particles that would give
rise to single particle decays induce a friction-like term, which in
turn gives rise to dissipational dynamics for the zero mode. This dissipation
allows for the energy contained in the zero mode's oscillations to be used for
the creation of light quanta which will eventually thermalize and allow the
standard big bang picture to proceed\cite{dolgov}.

This scenario has been questioned recently when it was realized that the
evolution of the zero mode of the inflaton induces large amplification of the
quantum fluctuations resulting in copious particle
production\cite{lindekov,brand,rehe}. We call this mechanism of particle
production {\it induced amplification}.  (In the case of exactly periodic
fields this reduces to parametric resonance \cite{dau}). When the zero mode of
the inflaton field has a large amplitude, this mechanism of induced
amplification is much more efficient for particle production than single
particle decay. In this letter we compare the time scales for damping and
particle production for the mechanisms of induced amplification and single
particle decay as well as the distribution of particles produced. It is
important to understand the different processes, because these time scales
determine the reheating temperature\cite{linde}.

We consider the situation in which the inflaton is coupled to lighter scalar
and fermion fields and track the behavior of the expectation value the inflaton
zero mode, as well as the various modes of the lighter field.  We find that the
modes of the lighter field are {\it not} distributed thermally, but are more
skewed towards modes with momentum $k \leq M_{\Phi}$. This may have
implications for the thermalization phase of the reheating
process. Interactions will tend to thermalize this spectrum to temperatures $
\approx M_{\Phi}$. The case of coupling to fermions is qualitatively different
>from the coupling to bosons because of Pauli blocking.

In both cases, we consider relaxation both in the linear regime, where the
amplitude of the zero mode is not large and an amplitude expansion of the
effective equation of motion can be used reliably, as well as non-linear
relaxation, where the initial amplitude is large; this is typically the
behavior of the inflaton in chaotic inflationary models\cite{chaotic}.

We use the techniques that we developed previously\cite{rehe} to
construct the effective equations of motion for the zero mode in the presence
of fluctuations due to the higher momentum modes. The reader is referred to
this and forthcoming work\cite{largepaper} for more details.

Throughout we will assume that the typical rates involved in
the particle production process are much larger than the expansion
rate of the
Universe $H$; this allows us to restrict ourselves to the Minkowski space
case.

\noindent{{\bf Linear Relaxation of the Inflaton:}}
Consider the following Lagrangian for the inflation $\Phi$ a scalar
$\sigma$ and a
fermion field $\psi$:
\begin{eqnarray}
{\cal L} & = & \frac{1}{2} \; (\partial \Phi)^2 + \frac{1}{2}  \;
(\partial \sigma)^2
- V(\Phi, \sigma)\nonumber +
{\bar{\psi}} (i {\not\!{\partial}} -M_{\psi}
 -y  \; \Phi  ) \psi\\
V(\Phi, \sigma) & = & \frac{1}{2} \; M_{\Phi}^2  \; \Phi^2 +
\frac{\lambda_{\Phi}}{4!} \; \Phi^4 + \frac{1}{2}  \; M_{\sigma}^2  \;
\sigma^2 +
\frac{\lambda_{\sigma}}{4!} \; \sigma^4 + \frac{g}{2}  \; \sigma^2  \; \Phi^2,
\label{yukalan}
\end{eqnarray}
where $\sigma$, $\psi$ are the lighter
degrees of freedom ($M_\Phi>>M_\sigma,  M_\psi$). As in our previous
work\cite{rehe} we write $\Phi(\vec{x},t)=\delta(t)+\chi(\vec{x},t)$
where $\delta(t)$ is the expectation value of the zero mode of $\Phi$
in the non-equilibrium density matrix and
$\langle \chi(\vec{x},t) \rangle = 0$\cite{rehe}. The expectation
value of the $\sigma$ field is assumed zero; this is consistent with
the equations of motion.
Following the formalism of \cite{rehe}, we arrive at the {\it linearized},
non-equilibrium equation of motion for $\delta(t)$:

\begin{equation}
\ddot{\delta}(t)+M_{\Phi}^2  \; \delta(t)+\int_{-\infty}^t
\delta(t')  \; \Sigma_r(t-t')\, dt' = 0. \label{ampeq}
\end{equation}
The kernel $\Sigma_r(t-t')$ is obtained in the loop expansion and is
determined by the {\it retarded} self-energy.
This self-energy needs appropriate substractions which are absorbed
into mass and wave function renormalization. From now on, all masses
and couplings will be the renormalized ones. We choose to start the
evolution at $t=0$ using the initial conditions
$\delta(0) =\delta_0, \; \dot{\delta}(0)=0$. This equation can be solved via
the Laplace transform\cite{rehe,largepaper}. The inverse Laplace transform
is performed along the Bromwich contour and requires the identification of
the singularities of the inverse propagator:
$G(s)=\left[s^2+M^2_{\Phi}+\Sigma_r(s)\right]^{-1}$ where $s$ is the
Laplace transform variable. We have to distinguish different cases:
{\bf i) Unbroken symmetry case:} a) If $y=0$, the  lowest order
contribution to $\Sigma_r(t-t')$ is given by the
two-loop retarded self-energy which has a cut  for $s^2 > -(M_{\Phi}+
2M_{\sigma})^2$.

For weak coupling $G(s)$ has a one-particle pole below the three particle
threshold. The contribution of this pole gives oscillatory behavior of
$\delta(t)$ with the period determined by the position of the pole.  The
contribution from the continuum to the long time dynamics is determined by the
behavior of the imaginary part of $\Sigma_r$ near threshold and it yields a
power law $t^{-3}$ at long times for all non-zero masses. The imaginary part of
the self-energy does not indicate exponential relaxation and although the
amplitude of the oscillation damps out, it reaches a non-zero asymptotic limit
at long times. It is straightforward to prove that the ratio of the amplitudes
at two different times is a renormalization group invariant\cite{largepaper}.
b) If $y\neq 0$ then the kernel receives a contribution at one loop from the
fermion-antifermion intermediate state. The self-energy has a cut beginning at
$s^2 = -4M^2_{\psi}$. If $M_{\Phi} > 2M_{\psi}$, the one-particle pole moves
off the physical sheet into the second sheet and for weak coupling, the
spectral density has the Breit-Wigner form with a narrow resonance at the
renormalized value of the scalar mass and width $\Gamma \approx {y^2}
M_{\Phi}\slash{8\pi}$ (for $M_{\Phi} \gg M_{\psi}$). The expectation value
$\delta(t)$ now relaxes exponentially with a relaxation time $t_{rel} =
\Gamma^{-1}$, for $t< \Gamma^{-1}\ln(M_{\Phi}/\Gamma)$, but eventually the long
time behavior is dominated by the behavior of the discontinuity of the
imaginary part of the self energy near threshold leading to a $t^{-5/2}$ power
law behavior. Two subtractions of the self-energy are needed, resulting in mass
and wave function renormalization. If $M_{\Phi} < 2M_{\psi}$ there is a
one-particle pole below the two-particle threshold i.e. in the physical sheet,
and the long time behavior of $\delta(t)$ is oscillatory with nonzero
asymptotic amplitude. The continuum contribution falls off with the power law
above. Clearly an imaginary part for the self energy does not translate to
exponential relaxation.  {\bf ii) Spontaneously broken symmetry case:} If
$M^2_{\Phi}=-\mu^2 <0$, the new minimum is at $\Phi_0=\sqrt{6\mu^2 /
\lambda_{\Phi}}$, and we write
$\Phi(\vec{x},t)=\Phi_0+\delta(t)+\chi(\vec{x},t)$.  The self energy now has
two-particle cuts beginning at $s^2=
-4(M_{\sigma}+g\Phi_{0})^2,-4(M_{\psi}+y\Phi_{0} )^2, -8\mu^2$.  If $2\mu^2$
is greater than the lowest two particle threshold the one-particle pole moves
into the second, unphysical, sheet and the spectral density features a narrow
resonance, for weak coupling, at the renormalized value of the scalar mass of
total width given by $\Gamma_{tot}= \Gamma_{\psi}+\Gamma_{\sigma} \approx
{y^2}\sqrt{2}\mu\slash{8\pi}+{3g^2 \mu}\slash (4\pi\sqrt{2}
\lambda_{\Phi})$. In this case $\delta(t)$ exhibits exponential relaxation on a
time scale that is typically as the one above. Eventually, however, a power law
tail behaving as $t^{-3/2}$, which is determined by the two-particle scalar
threshold, will develop.

{F}rom our discussion above, we find two important points:
(i) In the case when the inflaton mass is above the multiparticle
thresholds, the narrow resonance gives rise to exponential relaxation for a
long period of time. During this time, the evolution is similar to that
obtained from the effective evolution equation $\ddot{\delta}+m^2 \; \delta
+\Gamma_{tot}  \;\dot{\delta}=0$ with $\Gamma_{tot}$ given by the
total decay rate. The
equation of motion (\ref{ampeq}) provides a ``proof'' of this effective
description; however this only applies {\it for small amplitudes}. (ii)
Usually the imaginary part of the self energy is identified with a
``thermalization'' or relaxation rate. We have seen above that this is only the
case whenever there is a resonance i.e. a pole in the unphysical sheet. If
there is a one particle pole below the multiparticle threshold, then the long
time behavior is undamped and oscillatory. The continuum contribution falls off
with a power law which is determined completely by the behavior of the spectral
density at threshold.

\noindent{{\bf Non-Linear Relaxation of the Inflaton:}}
We now contrast the linear relaxation scenario with the non-linear one, which
is found by considering evolution equation for the expectation value of the
scalar field {\it including the back-reaction of the quantum fluctuations} by
absorbing $\delta(t)$ in the definition of the now {\it time dependent} mass
term for the scalars and fermions:

\begin{equation}
m^2_{\Phi}(t)= M^2_{\Phi}+\frac{\lambda_{\Phi}}{2} \;\delta^2(t) \; , \;
m^2_{\sigma}(t)= M^2_{\sigma}+g \;\delta^2(t) \; ,\;
m_{\psi}(t)= M_{\psi}+y \;\delta(t). \label{massesoft}
\end{equation}

We can then obtain the one-loop evolution equations\cite{rehe,largepaper}
\begin{equation}
\ddot{\delta}+M^2_{\Phi} \;\delta+\frac{\lambda_{\Phi}}{6} \;\delta^3+
\delta \;\frac{\lambda_{\Phi}}{6} \;\langle \chi^2(\vec{x},t)
\rangle+g \;\delta  \;
\langle \sigma^2(\vec{x},t) \rangle+y  \;\langle {\bar{\psi}}(\vec{x},t)
\psi(\vec{x},t) \rangle =0, \label{nonlineq}
\end{equation}
where the non-equilibrium expectation values are the one-loop Feynman graphs
but with the time dependent masses given by eq.(\ref{massesoft}):

\begin{eqnarray}
&& \langle \chi^2(\vec{x},t) \rangle    =  \int \frac{d^3k}{(2\pi)^3}
 \frac{|U^2_k(t)|}{2\omega_{\Phi}(k)} \; \; , \; \;
\langle \sigma^2(\vec{x},t) \rangle  =  \int \frac{d^3k}{(2\pi)^3}
\frac{|V^2_k(t)|}{2\omega_{\sigma}(k)} \label{sigmamode} \\
&& \langle {\bar{\psi}}(\vec{x},t) \psi(\vec{x},t) \rangle
   =  -2 \int  {d^3 k \over (2\pi)^3 } \left[ 1-2k^2 f^{\dagger}_{k}(t) f_{k}
(t)
\right]. \label{fermimode}
\end{eqnarray}
The mode functions  $U_k(t) , \;  V_k(t), \; f_k(t)$
obey\cite{largepaper}

\begin{eqnarray}
&& \left[\frac{d^2}{dt^2}+k^2+m^2_{\Phi}(t)\right]U_k(t)  =  0
\; \; , \; \;  U_k(0) = 1 \; , \; \dot{U}_k(0) =-i\omega_{\Phi}(k)
\label{fimodeq}\\
&& \left[\frac{d^2}{dt^2}+k^2+m^2_{\sigma}(t)\right]V_k(t)  =
0  \; \; , \; \;  V_k(0) = 1 \; , \; \dot{V}_k(0)
=-i\omega_{\sigma}(k) \label{sigmodeq} \\
&& \left[ \frac{d^2}{dt^2} + k^2 +m^2_{\psi}(t)-i {\dot{m}_{\psi}}(t)
\right] f_{k} (t)                         =  0
\; \; , \; \;  f_k(0) =\frac{1}{\sqrt{\omega_{\psi}(k)
(\omega_{\psi}(k)+m_{\psi}(0))}}
\label{fermodes}\\
&& {\dot{f}}_k(0)
     =-i \omega_{\psi}(k)f_k(0)  \; , \quad   \omega_{a}(k) =
\sqrt{k^2+m^2_{a}(0)} \; \; , \; \;
a=\Phi,\psi,\sigma .
\end{eqnarray}
The mode functions $f_{k} (t)$ define the spinor solutions of the
homogeneous Dirac equation.

{F}rom eqs.(\ref{fimodeq} - \ref{fermodes}), induced
amplification of quantum fluctuations becomes evident through the time
dependent mass; the mode functions obey harmonic oscillator equations with a
time-dependent frequency. If the time dependence were truly periodic this would
be parametric amplification through the presence of resonances and
forbidden bands\cite{lindekov,brand,dau}. However, when back reaction is
incorporated as
in the set of equations above, these time dependent frequencies are not exactly
periodic precisely because of the damping effects. This case should be
distinguished from parametric amplification and we refer to this case as
induced amplification.

The first order quantum correction to the evolution equation is obtained by
replacing the classical trajectories for $\delta(t)$ in the mode equations
(\ref{fimodeq} - \ref{fermodes}). However, this will lead to
infinite particle production via parametric amplification. In
reference\cite{lindekov}, the classical evolution for $\delta(t)$ was kept in
the mode equations. We, however, will keep the {\it full evolution} for
$\delta(t)$ thus obtaining a {\it self-consistent one-loop approximation} that
incorporates the back-reaction of the field into the fluctuations and
vice-versa. In effect, this procedure is a non-perturbative resummation of  a
large set of diagrams self-consistently.

Induced amplification will give rise to damped evolution for
$\delta(t)$, whose amplitude diminishes at long times. Thus, incorporating the
{\it full} evolution of $\delta(t)$ in the mode equations will shut-off
particle production at long times as the energy of the scalar field is
transferred to the produced particles.  This closed system of consistent
equations will ensure that particle production will be finite.

To obtain the dynamics of the zero mode, we will integrate the equations above
numerically, using the initial conditions $\delta(0)=\delta_0, \;
\dot{\delta}(0)= 0$. These conditions imply that the modes written above
correspond to positive frequency modes for $t<0$.
We have to renormalize the equations to one-loop
order absorbing  the relevant divergences  into mass, wave-function and
coupling constant renormalizations\cite{largepaper}. The final form for the
zero mode equation is as given by eq.(\ref{nonlineq}) in terms of
renormalized couplings and masses and the expectation values  have
been twice subtracted
with the aid of a high momentum cutoff $\Lambda$ (typically, $\Lambda \sim 100
|M_{\Phi}|$). Such a cutoff is necessary in any case in order to perform the
momentum integrals appearing in the expectation values numerically. The results
have been checked to insure that there is no cutoff sensitivity in
physical quantities. The higher mode equations take the
same form as in eqs.(\ref{fimodeq} - \ref{fermodes}), but in term
of the renormalized parameters\cite{rehe,largepaper}.

We now use these equations, written in terms of the dimensionless quantities
$\eta(\tau)=\Phi(t) \sqrt{\lambda_{\Phi}/ 6 M^2_{\Phi}};~~ \tau =
M_{\Phi} \; t$, to anyalze the  dynamics of the zero mode as well
as to compute the number of $\sigma,~~ \psi$ quanta produced by induced
amplification\cite{rehe,largepaper}.
{\bf iii) Unbroken symmetry case:} a) $y=0$:  Figure 1 shows
$\eta(\tau)$ vs. $\tau $
for ${\lambda_{\Phi}}\slash{8\pi^2}=0.2,~~ g=\lambda_{\Phi},~~
M_{\sigma}=0.2 \; M_{\Phi}, ~~ \eta(0)=1.0,~~ \dot{\eta}(0)=0$. The amplitude
damps out dramatically within a few periods of oscillation. The asymptotic
behavior corresponds to small undamped oscillations. This is to be contrasted
with the linear relaxation case, in which the two-loop self-energy leads to
undamped asymptotic oscillations and a power law behavior at long
times. We also find that the number of particles in a volume $M^{-3}_{\Phi}$,
as a function of wave number $k$, is peaked at $k \leq M_{\Phi}$. Induced
amplification is {\it much more} effective than one-particle decay in
producing low momentum particles and drawing energy from the
background field, thus damping its evolution. One-particle decay
considerations would predict a slow power law
tails to a final amplitude that is perturbatively close (notice that
the perturbative parameter is ${\lambda_{\Phi}}\slash{8\pi^2}$) to the
initial.

We also find\cite{largepaper} that both the quantum fluctuation $\langle
\sigma^2(\vec{x},t) \rangle$ and the total number of produced particles as a
function of time grow rapidly within the time scale in which damping is most
dramatic, as can be seen in figure (1), clearly showing that damping is a
consequence of induced amplification of fluctuations and particle production.
b) $y\neq 0,\ g=0$: Figure 2 shows $\eta(\tau)$ vs $\tau$ for $M_{\psi}=0
,~~y^2/ \pi^2=0.5,~~\lambda_{\Phi}/6y^2=1.0$. Clearly there is some initial
damping that shuts off early on.  This is in marked contrast with the
prediction of the linear relaxation approximation that predicts exponential
relaxation with a relaxation time (in the above units) $\tau_{rel} \approx 0.1$
for the above value of the couplings. We find\cite{largepaper} that the number
of particles in a volume $M^{-3}_{\Phi}$ as a function of wave number is peaked
at $k \leq M_{\Phi}$ with maximum value of 2 given by the exclusion
principle. We interpret the lack of damping as a consequence of the Pauli
blocking; all the low momentum states are fully occupied. In the initial stages
of evolution an excited state quickly ensues and the available fermionic states
are occupied blocking any further damping. Both the quantum fluctuation
$\langle \bar{\psi}(\vec{x},t)\psi(\vec{x},t) \rangle$ and the number of
produced particles as a function of time grow very rapidly within the time
scale of the initial damping, but then oscillate around a constant value at the
time when the damping effects have shut off\cite{largepaper}.  {\bf iv)
Spontaneously broken symmetry case:} $y = 0$.  Figure 3 shows the time
evolution of $\eta(\tau)$ vs $\tau$ for ${\lambda_{\Phi}}\slash{8\pi^2}=0.2,~~
g/ \lambda_{\Phi} =0.05,~~ M_{\sigma}=0.2 \; M_{\Phi},~~ \eta(0)=0.6,~~
\dot{\eta}(0)=0$. For these values of the parameters, linear relaxation
predicts exponential damping with a relaxation time $\tau_{rel} \approx
360$. This relaxation time is about {\it six to eight times larger} than that
revealed in figure (3). Furthermore, this figure clearly shows that the
relaxation is {\it not} exponential. The asymptotic value of $\eta(\tau)$
oscillates around the minimum of the effective action which, as observed in the
figure, has been displaced by a large amount from the tree level value $\eta
=1$. We notice numerically a slow damping in this asymptotic behavior, which
could be related to the linear relaxation time, though we were unable to
establish this numerically. We find\cite{largepaper} that the number of
produced particles in a volume $M_{\Phi}^{-3}$ is peaked at low wave number
with a fairly large contribution within the region $k \approx M_{\Phi}$. As a
function of time the quantum fluctuation $\langle \sigma^2(\vec{x},t) \rangle$
grows rapidly within the time scale in which damping is most pronounced and the
same behavior is observed for the number of particles created as a function of
time, thus leading to the conclusion that the rapid non-linear relaxation is
clearly a result of particle production and the growth of quantum
fluctuations. In all cases we find the total density of produced particles to
be ${\cal{N}} = N M_{\Phi}^3 ,~~ N \approx {\cal{O}}(1)$.

\noindent{{\bf Conclusions:}} We have learned that induced amplification of
quantum fluctuations in the non-linear regime is a highly efficient mechanism
for damping the evolution of the inflaton zero mode as well as for creating
particles. The typical time scales associated with this damping mechanism are
{\it much shorter} than the time scales associated with single particle
decays. This leads us to conclude that there may be a wide separation between
the time scales of induced particle production and that of thermalization of
these produced particles via collisional relaxation, since this latter scale is
typically related to perturbative couplings and thus fairly large. Single
particle decay will be operative on the same time scale as collisional
thermalization.

The final reheating temperature, however, will be determined by the total
number of particles produced. Typically, $T_{reh} \approx
{\cal{N}}^{1/3}$. Similarly, the total entropy produced will depend on the
number of produced particles.  Assuming that the process of single particle
decay produces far fewer particles than induced amplification and that a
thermalized state ensues from collisional relaxation, one can estimate the
final reheating temperature by evolving the equations above and following the
evolution of the number of particles. Clearly this number will depend on the
particular details such as coupling, masses but not in the perturbatively
obvious manner. Assuming that the produced particles thermalize via collisions
in the cases that we studied we typically obtain reheating temperatures
$T_{reh} \approx M_{\Phi}$\cite{largepaper} which is about two orders of
magnitude less than the estimate from the elementary theory of reheating
\cite{linde}.  However, it is not clear how long a distribution that is so
different from thermal will take to thermalize. In an expanding universe, this
time lag will serve to dilute the energy density of the produced
particles. Ultimately, a realistic calculation will involve the numerical
evolution of the self-consistent equations. Furthermore our study will have to
be extended to an FRW cosmology since the expansion of the universe can lead to
many effects not accounted for here. This work is in progress.

{\bf Acknowledgements:} D.B., H. J. de V. and R. H. would like to
thank  Andrei Linde for fruitful discussions. D. B. and D.-S. L. thank
F. Cooper and R. Willey for stimulating conversations, D.-S. L. thanks Bei Lok
Hu
for illuminating conversations and thank N.S.F for  partial
support  through Grant \#: PHY-9302534, INT-9216755.
R. H. and A. S. were partially supported by U.S.DOE under contract
DE-FG02-91-ER40682.

\newpage

{\bf Figure Captions}

Fig. 1: Unbroken symmetry case: $y=0,~~
{\lambda_{\Phi}}\slash{8\pi^2}=0.2,~~ g=\lambda_{\Phi},~~M_{\sigma}=0.2\,
M_{\Phi},~~\eta(0)=1.0 ,~~ \dot{\eta}(0)=0$.

Fig. 2: Unbroken symmetry case:
 $g=0,~~ M_{\psi}=0
,~~y^2/\pi^2=0.5,~~\lambda_{\Phi}/6y^2=1.0,~~\eta(0)=1.0,~~
\dot{\eta}(0)=0$.

Fig. 3: Broken symmetry case: $y=0,~~ {\lambda_{\Phi}}\slash{8\pi^2}=0.2,~~ g
/ \lambda_{\Phi} =0.05,~~ M_{\sigma}=0.2 \, M_{\Phi},~~ \eta(0)=0.6,~~
\dot{\eta}(0)=0$.


\begin{thebibliography}{99}
\bibitem{linde} A. D. Linde, Particle Physics and Inflationary
Cosmology, (Harwood, Chur,
Switzerland, 1990), ``Lectures on Inflationary Cosmology'' hep-th/9410082.
\bibitem{abbott}A. D.Dolgov and A. D. Linde, Phys. Lett. 116B, 329 (1982);
 L. F. Abbott, E. Farhi and M. B. Wise, Phys. Lett. 117 B, 29 (1982)
\bibitem{dolgov} A. D. Dolgov  and D. P. Kirilova,
Sov. J. Nucl. Phys. 51, 273 (1990);
J. Traschen and R. Brandenberger, Phys. Rev. D42, 2491 (1990).
\bibitem{lindekov} L. Kofman, A. Linde and A. A. Starobinsky,
Phys. Rev. Lett.73, 3195 (1994).
\bibitem{brand} ``Universe Reheating after Inflation'' by Y. Shtanov,
J. Traschen and R. Brandenberger, Brown University preprint BROWN-HET-957,
hep-ph/9407247.
\bibitem{rehe} D. Boyanovsky, H. J. de Vega, R. Holman, D.-S. Lee
and A. Singh, Phys. Rev. D51, 4419 (1995).
D. Boyanovsky, H. J. de Vega and R. Holman ,
Phys. Rev. D49, 2769 (1994).
\bibitem{chaotic} A.D. Linde, Phys. Lett. 129B, 177 (1983).
\bibitem{largepaper} D. Boyanovsky, M. D'Attanasio, H. J. de Vega,
R. Holman,  D.S.-Lee and A.S. Singh, in preparation.
\bibitem{dau} L. D. Landau and E. M. Lifshits, Mechanics, Pergamon
Press, London, 1958.



\end{thebibliography}
\end{document}